\documentstyle[prl,aps,multicol]{revtex}
\input{epsf}
\begin{document}
\draft
\title{Eigenstates of Operating Quantum Computer:\\ 
Hypersensitivity to Static Imperfections} 
\author{Giuliano Benenti$^{(a)}$, Giulio Casati$^{(a,b)}$, 
Simone Montangero$^{(a)}$, and Dima L. Shepelyansky$^{(c)}$}  
\address{$^{(a)}$International Center for the Study of Dynamical 
Systems, Universit\`a degli Studi dell'Insubria and} 
\address{Istituto Nazionale per la Fisica della Materia, 
Unit\`a di Como, Via Valleggio 11, 22100 Como, Italy}   
\address{$^{(b)}$Istituto Nazionale di Fisica Nucleare, 
Sezione di Milano, Via Celoria 16, 20133 Milano, Italy}   
\address{$^{(c)}$Laboratoire de Physique Quantique, UMR 5626 du CNRS,
Universit\'e Paul Sabatier, 31062 Toulouse Cedex 4, France}
\date{December 21, 2001}
\maketitle

\begin{abstract} 
We study the properties of eigenstates of an operating quantum computer 
which simulates  the dynamical evolution in the regime of quantum chaos.
Even if the quantum algorithm is polynomial in number of qubits $n_q$,
it is shown that the ideal eigenstates become mixed
and strongly modified by static imperfections above a certain
threshold which drops {\it exponentially} with $n_q$.
Above this threshold the quantum eigenstate entropy grows linearly with $n_q$
but the computation remains reliable 
during a time scale which is polynomial 
in the imperfection strength and in $n_q$.
\end{abstract} 
\pacs{PACS numbers: 03.67.Lx, 05.45.Mt, 24.10.Cn}  

\begin{multicols}{2}
\narrowtext

Feynman suggested that a quantum computer could simulate 
quantum mechanical systems exponentially faster than a 
classical computer \cite{feynman} while Shor significantly
extended this class by his ground-breaking algorithm for integer 
factorization \cite{shor}.
More recently, a few quantum algorithms which achieve 
the exponential speedup have been developed for 
various quantum and classical  physical systems, ranging from     
some many-body problems \cite{lloyd} 
to spin lattices \cite{optics}, and models of 
quantum and classical chaos \cite{schack,krot,simone}. 
These algorithms have been constructed for idealized perfect
quantum computers. In reality, any laboratory implementation of 
a quantum computer would involve imperfections \cite{static}. The first 
investigations have shown a certain stability of quantum evolution and
algorithms with respect to decoherence effects \cite{paz}, noisy gates 
\cite{zoller,zurek,song}, and static imperfections 
\cite{simone,static}. These studies have focused on the fidelity of 
quantum computation as a function of time during the 
realization of a given quantum algorithm.   

In this Letter, we study the properties of the eigenstates of 
an operating quantum computer in the presence of static imperfections. 
The computer is simulating efficiently the time evolution of a dynamical 
quantum system described by the sawtooth map \cite{simone}. We 
focus on the regime of quantum ergodicity, in which eigenfunctions 
are given by a complex superposition of a large number of quantum 
register states. In this regime, the effect of a perturbation is  
enhanced by a factor which is {\it exponential} in the number of 
qubits. 
This phenomenon has close links with the 
enormous enhancement of weak interactions in heavy nuclei
\cite{flambaum}.
In the following we illustrate 
this general effect for the case of static imperfections in a 
realistic model of quantum computer hardware.   

The classical sawtooth map is given by
\begin{equation}
\overline{n}={n}+k(\theta-\pi),
\quad
\overline{\theta}=\theta+T\overline{n},
\label{clmap}
\end{equation}
where $(n,\theta)$ are conjugated action-angle variables
($0\le \theta <2\pi$), and the bars denote the 
variables after one map step. Introducing the rescaled momentum 
variable $p=Tn$, one can see that the classical dynamics depends only on
the single parameter $K=kT$, so that the motion is stable for $-4<K<0$
and completely chaotic for $K<-4$ and $K>0$.
The quantum evolution for one map iteration is described
by a unitary operator $\hat{U}$ acting on the wave function
$\psi$:
\begin{equation}
\overline{\psi}=\hat{U}\psi =
e^{-iT\hat{n}^2/2}
e^{ik(\hat{\theta}-\pi)^2/2}\psi,  
\label{qumap}
\end{equation}
where $\hat{n}=-i\partial/\partial\theta$ (we set $\hbar=1$).
The classical limit corresponds to
$k\to \infty$, $T\to 0$, and 
$K=kT=\hbox{const}$. 
In this Letter, we study the quantum sawtooth map (\ref{qumap})  
in the regime of quantum ergodicity, with $K=\sqrt{2}$, 
$-\pi\leq p <\pi$ (torus geometry). 
The classical limit is obtained by increasing the number of qubits
$n_q=\log_2 N$ ($N$ number of levels), with $T=2\pi/N$ 
($k=K/T$, $-N/2\leq n < N/2$). 
The quantum algorithm  \cite{simone} simulates 
with exponential efficiency the quantum dynamics (\ref{qumap}) 
using a register of $n_q$ qubits. 
It is based on the forward/backward quantum Fourier transform 
\cite{qft} between the $\theta$ and $n$ representations and 
requires $2 n_q$ Hadamard gates and $3 n_q^2 -n_q$ 
controlled-phase-shift gates per map iteration \cite{simone}.   

Following \cite{static}, we model the quantum computer hardware 
as an one-dimensional array of qubits (spin 
halves) with static imperfections, i.e. fluctuations in the individual 
qubit energies and residual short-range inter-qubit couplings . 
The model is described by the many-body Hamiltonian  
\begin{equation}
\hat{H}_{\hbox{s}}=\sum_i (\Delta_0+\delta_i)\hat{\sigma}_i^z +
\sum_{i<j}J_{ij}\hat{\sigma}_i^x\hat{\sigma}_j^x,
\label{imperf}
\end{equation}
where the $\hat{\sigma}_i$ are the Pauli matrices for the qubit $i$,
and $\Delta_0$ is the average level spacing for one qubit. 
The second sum in (\ref{imperf}) runs over nearest-neighbor qubit 
pairs, and $\delta_i$, $J_{ij}$ are randomly and uniformly distributed
in the intervals $[-\delta/2,\delta/2]$ and $[-J,J]$,
respectively.
We study numerically the many-qubit eigenstates of the quantum 
computer (\ref{imperf}) running the quantum algorithm 
described above. 
The algorithm is realized by
a sequence of instantaneous and perfect one-
and two-qubit gates, separated by a time interval $\tau_g$,
during which the Hamiltonian (\ref{imperf}) gives unwanted
phase rotations and qubit couplings. 
We assume that the average phase accumulation given by 
$\Delta_0$ can be eliminated, e.g. by means of refocusing 
techniques \cite{chuang}.  

Since the evolution operator (\ref{qumap}) remains periodic 
in the presence of static imperfections, 
$\hat{U}^{(\epsilon)}(\tau+T)=\hat{U}^{(\epsilon)}(\tau)$ 
($\epsilon\equiv \delta \tau_g$), 
all the informations about the system dynamics are included in 
the quasienergy eigenvalues 
$\lambda_\alpha^{(\epsilon)}$  
and eigenstates 
$\phi_\alpha^{(\epsilon)}$ 
of the Floquet operator: 
\begin{equation} 
\hat{U}^{(\epsilon)}(T)  
\phi_\alpha^{(\epsilon)}=\exp(i \lambda_\alpha^{(\epsilon)}) 
\phi_\alpha^{(\epsilon)}.  
\label{floquet} 
\end{equation} 

\begin{figure} 
\centerline{\epsfxsize=4.2cm\epsffile{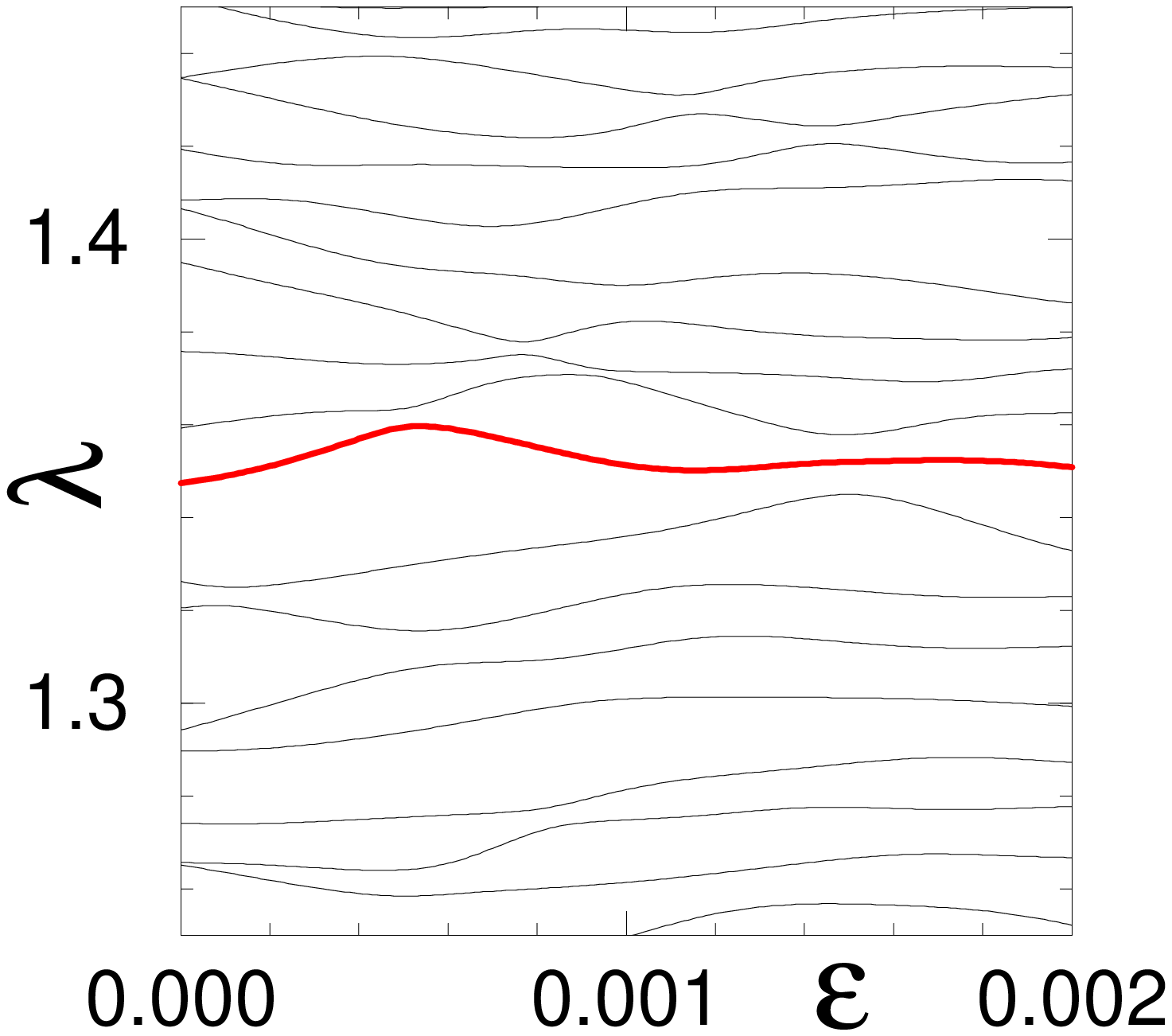}
\hfill\epsfxsize=4.2cm\epsffile{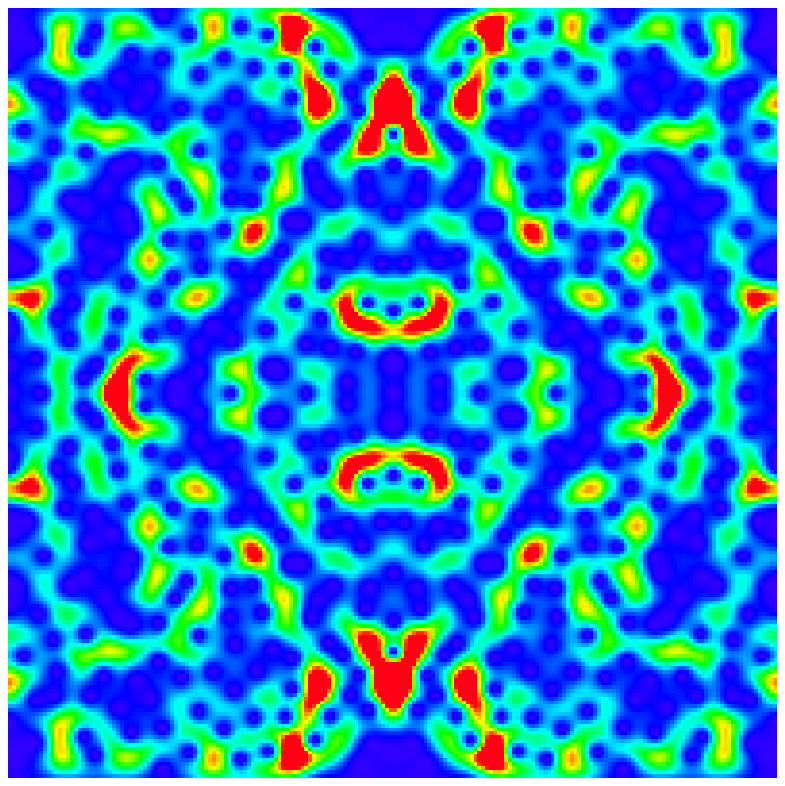}}  
\centerline{\epsfxsize=4.2cm\epsffile{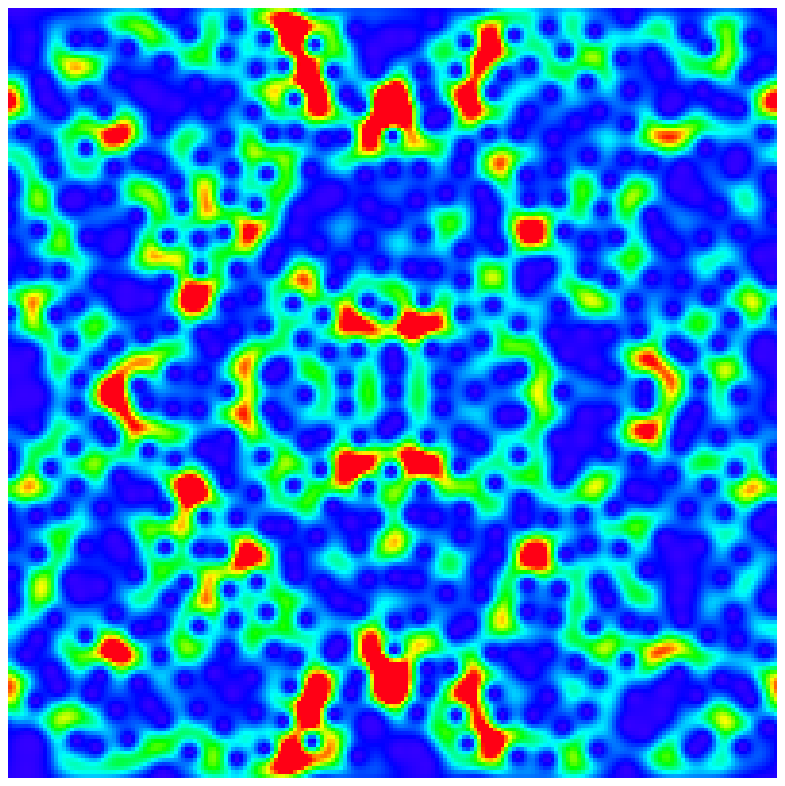}
\hfill\epsfxsize=4.2cm\epsffile{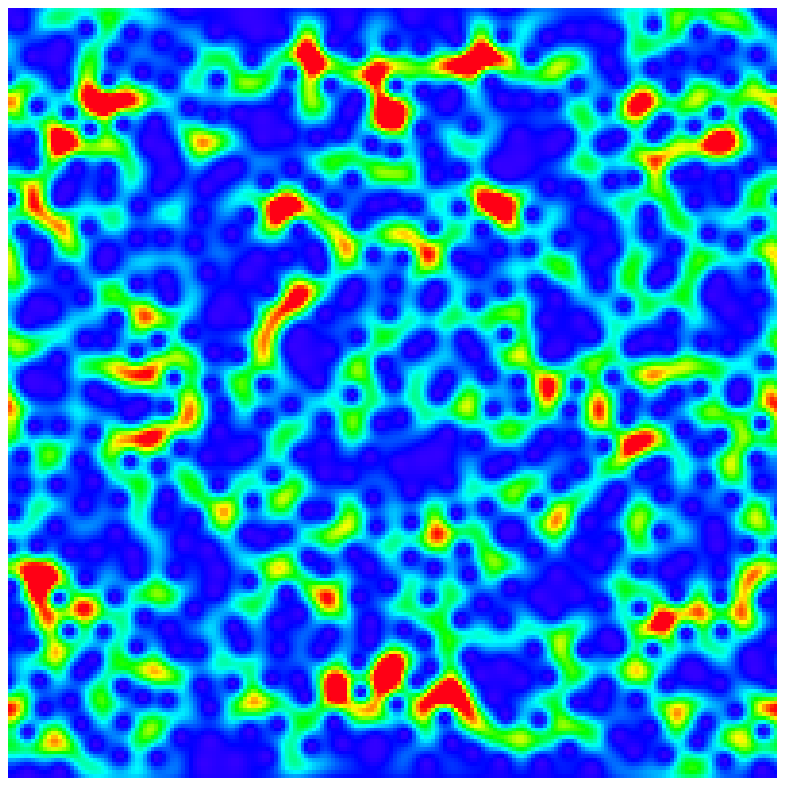}}  
\vspace{.2cm}
\caption{Parametric dependence of the quasienergy eigenvalues 
on the imperfection strength $\epsilon$ for a given random realization 
of $\delta_i$, at $J=0$, $n_q=9$ (top left);   
for the red-colored level the corresponding Husimi functions in
action-angle variables $(p,\theta)$ 
($-\pi \leq p < \pi$ -vertical axis- and 
$0\leq \theta < 2\pi$ -horizontal axis-) 
are given at $\epsilon=0$ (top right), 
$\epsilon=4\times 10^{-4}$ (bottom left), and $\epsilon=10^{-3}$ 
(bottom right).    
We choose the ratio of the action-angle uncertainties 
$s=\Delta p/\Delta\theta=T\Delta n/\Delta\theta=1$ 
($\Delta p \Delta\theta= T /2$).
The color is proportional to the density: blue for zero and 
red for maximal density. 
}
\label{fig1}       
\end{figure}

In Fig.\ref{fig1} (top left) we show the parametric dependence 
of the quasienergy eigenvalues on the dimensionless 
imperfection strength $\epsilon$, for a given   
realization of $\delta_i$, at $J=0$ and for $n_q=9$ qubits. 
One can clearly see the presence of avoided crossings, 
a typical signature of ergodic dynamics. 
The variation of a given quasienergy eigenstate with
$\epsilon$ is illustrated by the Husimi functions \cite{husimi} 
of Fig.\ref{fig1}. At $\epsilon=0$ the eigenfunctions display 
a complex pattern delocalized in the phase space 
(see Fig.\ref{fig1} top right). The symmetries of the Husimi 
functions ($\theta\to 2\pi-\theta$, $p\to -p$) are destroyed 
when $\epsilon\ne 0$ \cite{symmetries}. 
The eigenfunctions in the presence 
of imperfections give a good representation of the unperturbed  
($\epsilon=0$) eigenfunctions of the quantum sawtooth map 
(\ref{qumap}) at most until the first avoided crossing. 
After that one cannot make a one to one correspondence 
with the unperturbed case. This is confirmed by the last 
two Husimi functions of Fig.\ref{fig1}, taken for the 
chosen level in the vicinity of the first avoided crossing  
(bottom left, $\epsilon=4\times 10^{-4}$) and for a stronger 
imperfection strength (bottom right, $\epsilon=10^{-3}$). 
In the first case there is still some similarity                 
with the corresponding  Husimi function at $\epsilon=0$,  
while in the latter case any resemblance has been effaced.   

\begin{figure} 
\centerline{\epsfxsize=8.cm\epsffile{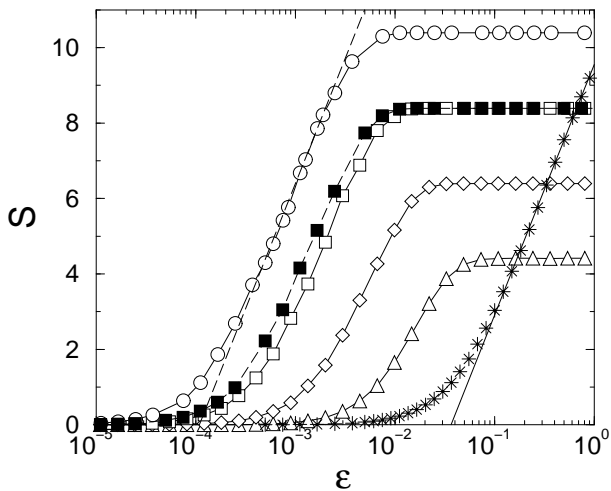}}
\caption{Quantum eigenstate entropy $S$ as a function of  
the scaled imperfection strength $\epsilon$ for $J=0$, 
$n_q=5$ (triangles), $7$ (diamonds), $9$ (empty squares), and 
$11$ (circles), for $J=\delta$, $n_q=9$ (filled squares), 
and for the single imperfection model at $n_q=11$ (stars). 
The straight lines give the theoretical estimates 
$2^S=A \epsilon^2 N$ (solid line) and  
$2^S=B \epsilon^2 n_q^5 N$ (taken at $n_q=11$, 
dashed line), with the numerically determined constants 
$A=0.37$ and $B=0.25$.}  
\label{fig2}
\end{figure}

\begin{figure} 
\centerline{\epsfxsize=8.cm\epsffile{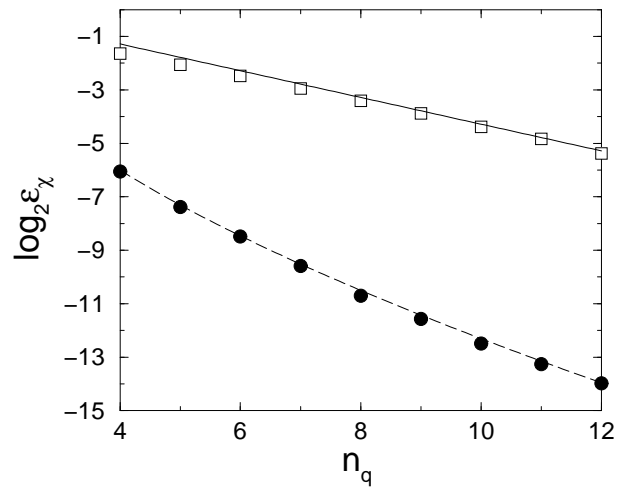}}
\caption{Dependence of the coupling $\epsilon_\chi$ at 
which $S=1$ on the number of qubits, for $J=0$ (circles) 
and for the single imperfection model (squares). 
The lines give the theoretical dependences  
$\epsilon_\chi = A^{-1/2} N^{-1/2}$ (above) and      
$\epsilon_\chi = B^{-1/2} N^{-1/2} n_q^{-5/2}$ (below), 
with the constants $A$ and $B$ obtained from the data 
of Fig.\ref{fig2}.} 
\label{fig3}
\end{figure}

A more quantitative indication of the similarity between    
exact and perturbed eigenstates is provided by   
the quantum eigenstate entropy,   
\begin{equation} 
S_\alpha=-\sum_{\beta=1}^N p_{\alpha\beta}\log_2 
p_{\alpha\beta},  
\end{equation} 
where $p_{\alpha\beta}=|\langle\phi_\beta^{(0)}|
\phi_\alpha^{(\epsilon)}\rangle|^2$. In this way $S_\alpha=0$ 
if $\phi_\alpha^{(\epsilon)}$ coincides with one eigenstate 
at $\epsilon=0$, $S_\alpha=1$ if $\phi_\alpha^{(\epsilon)}$ 
is equally composed of two ideal ($\epsilon=0$) eigenstates, and the 
maximal value $S_\alpha=n_q$ is obtained if all 
$\phi_\beta^{(0)}$ $(\beta=1,...,N=2^{n_q})$ contribute 
equally to $\phi_\alpha^{(\epsilon)}$. 
In order to reduce statistical fluctuations, we average 
$S_\alpha$ over $\alpha=1,...,N$ and over $3\leq N_D \leq 10^3$
random realizations 
of $\delta_i,J_{ij}$. In this way the 
total number of eigenstates is $N_D N\approx 10^4$. 
The variation of the average quantum entropy $S$ with 
$\epsilon$ is shown in Fig.\ref{fig2}. It demonstrates that 
$S$ grows from $S=0$ at $\epsilon=0$ to a saturation value 
$S\approx n_q$ corresponding to maximal mixing of unperturbed 
eigenstates. 
We study this crossover for $4\leq n_q \leq 12$ at $J=0$ and 
find that the mixing takes place at smaller values when 
$n_q$ increases. 
In Fig.\ref{fig3} we show that the critical imperfection strength 
$\epsilon_\chi$ at which $S=1$ drops exponentially with the 
number of qubits. 
This exponential dependence holds also in a simple 
toy model with a single impurity, in which an energy fluctuation $\delta$ 
is switched on for a single qubit and only for one time interval 
$\tau_g$ between two elementary gates (e.g., after the first 
quantum Fourier transform). 

\begin{figure} 
\centerline{\epsfxsize=8.cm\epsffile{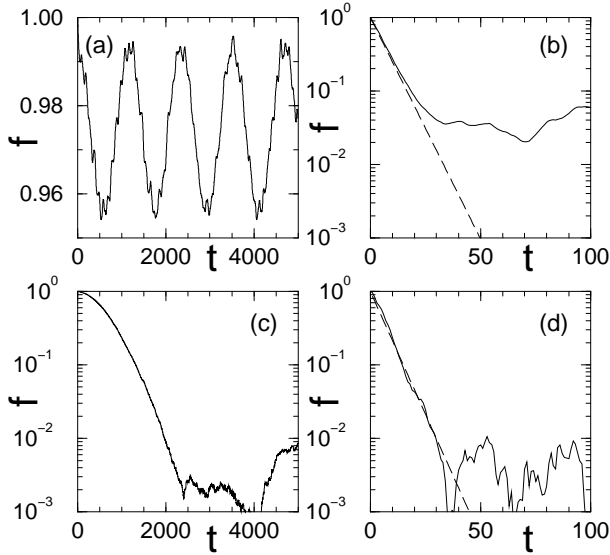}}
\caption{Fidelity as a function of time, for $n_q=9$ qubits, 
$J=0$, $\epsilon=10^{-4}$ (left) and $\epsilon=3\times 10^{-3}$ 
(right), with initial wave function a Floquet eigenstate 
at $\epsilon=0$ (top) or a momentum eigenstate (bottom). 
The dashed lines show the exponential 
decay $f(t)=\exp(-t/t_f)$, with $t_f\approx 7$.}
\label{fig4}
\end{figure}

\vglue -0.3cm
The exponential drop of the threshold $\epsilon_\chi$ can 
be understood following a theory originally developed 
for the parity breaking induced by weak interaction in the scattering of 
polarized neutrons on complex nuclei \cite{flambaum}.       
Indeed, due to quantum chaos in (\ref{qumap}), 
the Floquet problem (\ref{floquet}) has ergodic eigenstates, 
$\phi_\alpha^{(0)}=\sum_{m=1}^N c_\alpha^{(m)} u_m$, 
with $u_m$ being quantum register states and $c_\alpha^{(m)}$ 
randomly fluctuating components, with 
$|c_\alpha^{(m)}|\sim 1/\sqrt{N}$. 
For the model with a single imperfection
$\delta \hat{\sigma}_i^z$, acting on a 
time interval $\tau_g$, the transition matrix elements 
between unperturbed eigenstates have a  typical value: 
\begin{equation}
V_{\rm typ} \sim |\langle\phi_\beta^{(0)}| 
\delta\hat{\sigma}_i^z \tau_g | \phi_\alpha^{(0)}\rangle | =  
\epsilon |\sum_{m=1}^N c_\alpha^{(m)} c_\beta^{(m)\star}| 
\sim \epsilon / \sqrt{N}. 
\end{equation}  
\vglue -0.3cm
The last estimate for $V_{\rm typ}$ results from the 
sum of $N$ uncorrelated terms. 
Since the spacing between quasienergy eigenstates is 
$\Delta E \sim 1/N$, the threshold for the breaking of 
perturbation theory can be estimated as 
\begin{equation} 
V_{\rm typ}/\Delta E \sim \epsilon_\chi\sqrt{N} \sim 1. 
\end{equation} 
The analytical result $\epsilon_\chi\sim 1/\sqrt{N}$ 
is confirmed by the numerical data in Fig.\ref{fig3}. 
The same theoretical argument 
gives an exponential drop of $\epsilon_\chi$ for the static 
imperfection model (\ref{imperf}). 
In this case, the estimate can be obtained with 
$\delta\to\delta\sqrt{n_q}$ (sum of $n_q$ random detunings 
$\delta_i$) and $\tau_g\to\tau_g n_g\sim \tau_g n_q^2$. This gives 
$\epsilon_\chi\sim N^{-1/2} n_q^{-5/2}$, again in good 
agreement with the data of Fig.\ref{fig3}. For the case $J=\delta$, 
the threshold $\epsilon_\chi$ decreases by a factor $\approx 1.5$
at $n_q=9$ with respect to the $J=0$ case (see Fig.\ref{fig2}), 
since additional qubit couplings are introduced. 
We note that the hypersensitivity to perturbations has been proposed 
as a distinctive feature of chaotic dynamics \cite{caves}. 
However, the authors of Ref.\cite{caves} considered the effect 
of a stochastic environment, while we consider a closed Hamiltonian 
system.  
\vglue -0.3cm
In the mixing regime ($\epsilon>\epsilon_\chi$) the number $M$  
of unperturbed eigenfunctions $\phi_\beta^{(0)}$, which have a
significant projection over a given $\phi_\alpha^{(\epsilon)}$,
is exponentially large. For the single imperfection model one has
$M\sim 2^S\sim \Gamma/\Delta E\sim \epsilon^2 N$, since 
the mixing takes place inside a Breit-Wigner width given 
by the Fermi golden rule: $\Gamma\sim V_{\rm typ}^2/\Delta E 
\sim \epsilon^2$. The above estimate for $M$ is in agreement
with the numerical data of Fig.\ref{fig2} and is similar
to that one used in \cite{static,flambaum2} for onset of quantum chaos in 
the static model (\ref{imperf}).
We emphasize that this estimate implies that the quantum eigenstate  
entropy grows linearly with the number of qubits $n_q$. 
In the Fermi golden rule regime, the lifetime 
(measured in number of kicks) of an unperturbed eigenfunction is given by 
$\tau_\chi \sim 1/\Gamma \propto 1/\epsilon^2$ 
\cite{static,flambaum2}. 
If the imperfections are described by the model (\ref{imperf}), 
one has $\tau_\chi \sim 1/(\epsilon^2 n_q^5)$. 
Therefore a reliable quantum computing of the dynamical evolution of 
the model (\ref{qumap}) is possible up to a time scale which 
drops only {\it algebraically} with
\noindent
the number of qubits, in 
agreement with the findings of Ref. \cite{simone}. 
In Fig.\ref{fig4} we show the fidelity of quantum evolution, 
$f(t)=|\langle\psi^{(0)}(t)|\psi^{(\epsilon)}(t)\rangle|^2$. 
In the top figures, the initial state is an unperturbed eigenstate. 
For $\epsilon<\epsilon_\chi$ ($S<1$), the fidelity is very close 
to $1$ at all times, since the eigenstates are not mixed by 
the imperfections (see Fig.\ref{fig4}(a)). 
On the contrary, in the Fermi golden rule 
regime $\epsilon>>\epsilon_\chi$ a perturbed eigenstate, when 
decomposed into the unperturbed eigenstates, contains  
a large number of components ($S\gg 1$).
The distribution of these components over energy, 
called local density of states, has a 
typical Breit-Wigner shape of width $\Gamma$. 
Since its Fourier transform drives the fidelity 
decay \cite{static,flambaum2}, 
one obtains $f(t)\approx \exp (-\Gamma t)$,  
in agreement with the data of Fig.\ref{fig4}(b).
The exponential decay continues up to a value
$f\approx 1/2^S$ given by the inverse of the number 
of levels mixed inside the Breit-Wigner width. 
The case in which the initial wave function $\psi(0)$ is 
a momentum eigenstate is considered in Fig.\ref{fig4}(c,d).
In this case $\psi(0)$ 
projects significantly over order $N$ unperturbed eigenfunctions.  
Therefore, for $\epsilon<\epsilon_\chi$, $f(t)$ displays a 
Gaussian decay 
(see Fig.\ref{fig4}(c), and also Ref. \cite{simone}): in this regime
the imperfections do not change significantly the eigenstates
($S< 1$), but the initial state
is composed of many eigenstates and 
a Gaussian decay of $f(t)$ is espected from perturbation 
theory \cite{tomsovic,beenakker}.
Fig.\ref{fig4}(d) shows that in the Fermi golden rule regime 
($\epsilon>\epsilon_\chi$) the fidelity decays exponentially, 
with rate $\Gamma$ given by the Breit-Wigner width \cite{beenakker}. 
The decays stops when $f\approx 1/N$, namely when $f$ 
approaches the inverse of the dimension of the Hilbert space.  
In summary, we have shown that the eigenstates of a 
quantum computer simulating a  system with quantum chaos are 
hypersensitive to static imperfections: 
they are significantly different from the exact eigenfunctions 
above an imperfection strength threshold which drops 
exponentially with the number of qubits. 
This border is directly relevant for quantum algorithms which 
aim at computing  static properties of physical systems, 
for example energy eigenvalues and eigenvectors \cite{lloyd2}.    
We also note that quantum adiabatic algorithms \cite{farhi} 
assume the ability to drive the evolution of a quantum state 
towards the ground state of a specified Hamiltonian. 
Moreover, a few relevant quantum algorithms can be formulated 
in terms of determining the eigenvalues corresponding to 
eigenstates of a given unitary operator \cite{kitaev,ekert}. 
Further investigations are required to establish under which 
conditions the observed hypersensitivity of eigenstates 
to perturbations could be relevant for the stability of 
these algorithms.  

This work was supported in part by the EC RTN contract 
HPRN-CT-2000-0156, the NSF under grant No. PHY99-07949
 and (for D.L.S.) by the NSA and ARDA under 
ARO contract No. DAAD19-01-1-0553. Support from the PA 
INFM ``Quantum transport and classical chaos'' is 
gratefully acknowledged. 
\vglue -0.5cm

\end{multicols}


\begin{thebibliography}{99} 
\bibitem{feynman} R.P. Feynman, Int. J. Theor. Phys. 
{\bf 21}, 467 (1982). 
\bibitem{shor} P.W. Shor, in {\it Proceedings of the 35th Annual 
Symposium on Foundations of Computer Science}, edited by 
S. Goldwasser (IEEE Computer Society, Los Alamitos, CA, 1994), 
p. 124. 
\bibitem{lloyd} S.Lloyd, Science {\bf 273}, 1073 (1996). 
\bibitem{optics} A. S\o rensen and K. M\o lmer, Phys. Rev. Lett. 
{\bf 83}, 2274 (1999). 
\bibitem{schack} R. Schack, Phys. Rev. A {\bf 57}, 1634 (1998). 
\bibitem{krot} B. Georgeot and D.L. Shepelyansky, Phys. Rev. Lett. 
{\bf 86}, 2890 (2001); {\bf 86}, 5393 (2001).
\bibitem{simone} G. Benenti, G. Casati, S. Montangero, and 
D.L. Shepelyansky, Phys. Rev. Lett. 
{\bf 87}, 227901 (2001).  
\bibitem{static} B. Georgeot and D.L. Shepelyansky, 
Phys. Rev. E {\bf 62}, 3504 (2000); {\bf 62}, 6366 (2000); 
G. Benenti, G. Casati, and D.L. Shepelyansky,  
Eur. Phys. J. D {\bf 17}, 265 (2001). 
\bibitem{paz} C. Miquel, J.P. Paz, and R. Perazzo, Phys. Rev. A 
{\bf 54}, 2605 (1996). 
\bibitem{zoller} J.I. Cirac and P. Zoller, Phys. Rev. Lett. 
{\bf 74}, 4091 (1995).  
\bibitem{zurek} C. Miquel, J.P. Paz, and W.H. Zurek, 
Phys. Rev. Lett. {\bf 78}, 3971 (1997). 
\bibitem{song} P.H. Song and D.L. Shepelyansky, Phys. Rev. Lett. 
{\bf 86}, 2162 (2001).  
\bibitem{flambaum} O.P. Sushkov and V.V. Flambaum, Sov. Phys. Usp. 
{\bf 25}, 1 (1982). 
\bibitem{qft} See, e.g., A. Ekert and R. Jozsa, Rev. Mod. Phys.
{\bf 68}, 733 (1996).
\bibitem{chuang} See, e.g., N.A. Gershenfeld and I.L. Chuang, 
Science {\bf 275}, 350 (1997). 
\bibitem{husimi} The computation of Husimi functions is 
described in S.-J. Chang and K.-J. Shi, Phys. Rev. A 
{\bf 34}, 7 (1986).  
\bibitem{symmetries} In order to stress the perturbation induced 
symmetry breaking, in Fig.\ref{fig1} we consider the 
time-symmetric version of the map (\ref{qumap}):   
$\overline{\psi}=\hat{U}\psi =
e^{-iT\hat{n}^2/4}
e^{ik(\hat{\theta}-\pi)^2/2}
e^{-iT\hat{n}^2/4} \psi$. This map has  axial symmetry contrary
to central symmetry for map (\ref{qumap}).
\bibitem{caves} R. Schack and C.M. Caves, Phys. Rev. Lett 
{\bf 71}, 525 (1993); Phys. Rev. E {\bf 53}, 3257 (1996). 
\bibitem{flambaum2} V.V. Flambaum, Aust. J. Phys. {\bf 53}, 
489 (2000). 
\bibitem{tomsovic} N.R. Cerruti and S. Tomsovic, nlin.CD/0108016.
\bibitem{beenakker} Ph. Jacquod, P.G. Silvestrov, and 
C.W.J. Beenakker, Phys. Rev. E {\bf 64}, 055203(R) (2001).  
\bibitem{lloyd2} D.S. Abrams and S. Lloyd, Phys. Rev. Lett. 
{\bf 83}, 5162 (1999).  
\bibitem{farhi} E. Farhi, J. Goldstone, S. Gutmann, 
J. Lapan, A. Lundgren, and D. Preda, Science {\bf 292}, 472 (2001). 
\bibitem{kitaev} A.Yu. Kitaev, quant-ph/9511026.
\bibitem{ekert} R. Cleve, A.Ekert, C.Macchiavello, and M. Mosca, 
Proc. R. Soc. London A {\bf 454}, 339 (1998).    
\end{thebibliography}
\end{document}